\documentclass[preprint,preprintnumbers,superscriptaddress,amsmath]{revtex4}
\usepackage[dvips]{graphics}
\usepackage{epsfig}
\usepackage{psfrag}
\usepackage[psamsfonts]{amssymb}
\usepackage{amsmath}
\usepackage{indentfirst}
\usepackage{amssymb}
\usepackage{wrapfig}
\usepackage{url}

\begin{document}

\title{Generating Functions and Stability Study \\of Multivariate Self-Excited Epidemic Processes}
\author{A. Saichev}
\affiliation{Department of Management, Technology and Economics,
ETH Zurich, Kreuzplatz 5, CH-8032 Zurich, Switzerland}
\affiliation{Mathematical Department,
Nizhny Novgorod State University, Gagarin prosp. 23,
Nizhny Novgorod, 603950, Russia}
\email{saichev@hotmail.com,dsornette@ethz.ch}

\author{D. Sornette}
\affiliation{Department of Management, Technology and Economics,
ETH Zurich, Kreuzplatz 5, CH-8032 Zurich, Switzerland}

\begin{abstract}
We present a stability study of the class of multivariate self-excited Hawkes point processes,
that can model natural and social systems, including earthquakes, epileptic seizures
and the dynamics of neuron assemblies,
bursts of exchanges in social communities, interactions between Internet bloggers,
bank network fragility and cascading of failures, national sovereign default contagion, and so on.
We present the general theory of multivariate generating functions
to derive the number of events over all generations of various types that are triggered by a mother event
of a given type. We obtain the stability domains
of various systems, as a function of the topological structure of the mutual excitations
across different event types. We find that mutual triggering tends to provide
a significant extension of the stability (or subcritical) domain compared
with the case where event types are decoupled, that is, when an event of
a given type can only trigger events of the same type.
\end{abstract}

\date{\today}

\maketitle

\section{Introduction}

Many natural and social systems are punctuated by short-lived events that play a particularly
important role in their organization. Such events can be conveniently modeled mathematically by
so-called point processes \cite{Bremaud1,DVJ2007}. They are also called shot noise in physics
\cite{shotnoiseintro,Montroll12,Scher} or jump processes in finance and in economics \cite{ContTankov}.
These models are characterized by their (conditional) rate $\lambda(t | H_t)$ (also called ``conditional intensity'')
defined as the limit for small time intervals $\Delta$ of the probability that an event occurs between $t$ and
$t+\Delta$, given the whole past history $H_t$.  In mathematical notations, this reads
\begin{equation}
\lambda(t | H_t) = {\rm lim}_{\Delta \to 0} ~~{1 \over \Delta} {\rm Pr}({\rm event~ occurs~in}~ [t, t+\Delta] | H_t)~,
\end{equation}
where Pr$(X | H_t)$ represents the probability that event $X$ occurs, conditional
on the past history $H_t$. The symbol
$H_t$ represents the entire history up to time $t$, which includes all previous events.
This definition is straightforward to generalize for space-dependent intensities
 $\lambda(t,{\vec r} | H_t)$ and to include marks such as amplitudes or magnitudes (see below).
The standard Poisson memoryless process is the special case such that $\lambda(t | H_t)$ is constant,
i.e., independent of the past history.
Clustered point processes generalize the Poisson process by assuming that
the series of events are generated from a cluster center process,
which is often a renewal process, and a cluster member process.

The class of point processes that we study here was introduced by Hawkes in 1971
\cite{Hawkes1,Hawkes2,Hawkes3,Hawkes4}. It
is much richer and relevant to most natural
and social systems, because it describes ``self-excited'' processes.  This term means that the past events
have the ability to trigger future events, i.e., $\lambda(t | H_t)$ is a function of past events,
being therefore non-markovian.
Many works have been performed to characterize the statistical and dynamical properties
of this class of models, with applications ranging from geophysical \cite{Ogata88,Ogata1,Ogata2,Ogata3,HS02,SaiSor07},
medical \cite{SorOsorio10} to financial systems,
with applications to Value-at-risk modeling \cite{Chavezetal05}, high-frequency price processes
\cite{BauwensHautsch09}, portfolio credit risks \cite{Eymanetal10}, cascades
of corporate defaults \cite{Azizetal10}, financial contagion \cite{Aitsahaliaetal10},
and yield curve dynamics \cite{SalmonTham08}.

While surprisingly rich and  powerful in explaining empirical observations in a variety of
systems, most previous studies have used mono-variate self-excited point processes,
i.e., they have assumed the existence of only a single type of events, all the events presenting some
ability to trigger events of the same type.
However, in reality, in many systems,
events come in different types with possibly different properties,
while keeping a degree of mutual inter-excitations. Among others, this applies
to geo-tectonic deformations and earthquakes, to
neuronal excitations in the brain, to financial volatility bursts
in different assets, to defaults on debts in some firms or some industrial sectors,
to sovereign risks in some countries within a currency block,
to the heterogeneity of activity of bloggers on the Internet, and so on.

These observations suggest that multivariate self-excited point processes, which extend
the class of mono-variate self-excited point processes, provide a very important
class of models to describe the self-excitation (or  intra-triggering) as well as the mutual influences
or triggering between different types of events that occur in many natural and social systems.
These considerations have motivated us to present recently
the first exact analysis of some of the temporal properties of multivariate self-excited
Hawkes conditional Poisson processes \cite{SaiSorhierHaw11},
as they constitute powerful representations of a large variety
of systems with bursty events, for which past activity triggers future activity. The term ``multivariate''
refers here to the property that events come in different types, with possibly different intra and
inter-triggering abilities.  Ref.~\cite{SaiSorhierHaw11} was a first step towards
a systematic study of the multivariate self-excited point processes, first
mentioned by Hawkes himself in his first paper \cite{Hawkes1}, whose
full relevance has only been recently appreciated \cite{Zhang-Ma05,Aitsahaliaetal10}.

The present paper is a complementary study to our previous paper \cite{SaiSorhierHaw11},
which was focused on temporal properties, by studying the general stability conditions
of this class of models. Section 2 recalls the definition and notation of Hawkes processes,
starting from the monovariate version and extending to the general multivariate formulation.
Section 3 presents the formalism of multivariate generating functions
to derive the number of events over all generations of various types that are triggered by a mother event
of a given type. Section 4 gives the stability conditions using the mean numbers
of events of all generations. Subsection 4A provides the general relations.
Subsection 4B studies the case of symmetric mutual excitation abilities between events
of different types. Subsection 4C restricts to the case of just two different types of events,
that allows an in-depth analysis of the new features resulting from the inter-type excitations.
Subsection 4D presents the results obtained for a one-dimensional chain of directed
triggering in the space of event types. Subsection 4E generalizes subsection 4D by
studying a one-dimensional chain in the space of event types with nearest-neighbor
triggering. Subsection 4F presents a quantitative measure of the size of the 
subcritical domain that allows us to study the influence of the inter-type coupling strength.
Section 5 concludes.

\section{Definitions and notations for the multivariate Hawkes processes}

\subsection{Monovariate Hawkes processes}

Self-excited conditional Poisson processes
generalize the cluster models by allowing each event, including
cluster members, i.e., aftershocks, to trigger their own events according to some memory
kernel $h(t-t_i)$.
\begin{equation}
\lambda(t | H_t, \Theta) = \lambda_c(t) + \sum_{i | t_{i} < t}  h(t-t_i,)~,
\label{hyjuetg2tgj}
\end{equation}
where the history $H_t = \{ t_i \}_{1 \leq i \leq i_t,~ t_{i_t} \leq t < t_{i_t+1} }$ includes all events
that occurred before the present time $t$ and the sum
in expression (\ref{hyjuetg2tgj}) runs over all past triggered events.  The set of parameters is
denoted by the symbol $\Theta$.
The term $\lambda_c(t)$ means
that there are some external background sources occurring according to a Poisson process
with intensity $\lambda_c(t)$, which may be a function of time, but all other events can be both triggered
by previous events and can themselves trigger their offsprings.  This gives rise
to the existence of many generations of events.

Introducing ``marks'' or characteristics for each event leads to a first multidimensional
extension of the self-excited process (\ref{hyjuetg2tgj}). The generalization consists in associating
with each event some marks (possible multiple traits), drawn from some distribution $p(m)$,
usually chosen invariant as a function of time:
\begin{equation}
\lambda(t, M| H_t, \Theta) = p(M) \left( \lambda_c(t) + \sum_{i | t_{i} < t}  h(t-t_i, M_i) \right)~,
\label{hyjuetgq2h56h2tgj}
\end{equation}
where the mark $M_i$ of a given previous event now controls the shape and properties
of the triggering kernel describing the future offsprings of that event $i$.
The history now consists in the set of occurrence times of each triggered event and their
marks: $H_t = \{ t_i, M_i\}_{1 \leq i \leq N}$. The first factor $p(M)$ in the r.h.s. of
expression (\ref{hyjuetgq2h56h2tgj}) writes that the marks of triggered events
are drawn from the distribution $p(M)$, independently of their generation and waiting times.
This is a simplifying specification, which can be relaxed. Inclusion of spatial kernel
to describe how distance impacts triggering efficiency is straightforward.

From a theoretical point of view,
the Hawkes models with marks has been studied in essentially two directions:
 (i) statistical estimations of its parameters with corresponding residual analysis as goodness of fits \cite{Ozaki79,Ogate81lewis,OgataAkaike82,Ogata83estmi,Zhuangetal02,ZhuangOgata04,OgaZhu06,Zhuangthesis,Marsan08,SorUtkindeclus};
 (ii) statistical properties of its space-time dynamics \cite{HS02,SaiSor07,SorHelmsing02,Helmsordiffth,Helmsorgrasso,Helmsordirect03,HelmsorPredict03,Saichevsor05,SaiHelmSor05,SaiSorlifetime,SaiSorrenoetas,SorUtkinSai08}.

The advantage of the self-excited conditional Hawkes process
includes a very parsimonious description
of the complex spatio-temporal organization of systems characterized
by self-excitation of ``bursty'' events,
without the need to invoke ingredients other than the generally well-documented
stylized facts on the distribution of event sizes, the temporal ``Omori law''
for the waiting time before excitation of a new event and the productivity law controlling
the number of triggered events per initiator.

Self-excited models of point processes with additive structure of their intensity
on past events  \cite{Hawkes4} make them part
of the general family of branching processes \cite{Harris63}. The
crucial parameter is then the branching ratio $n$, defined as the mean number of events of first
generation triggered per event. Depending on applications, the branching ratio $n$ can vary with time,
from location to location and from type to type (as we shall see below
for the multivariate generalization). The branching ratio provides
a diagnostic of the susceptibility of the system to trigger activity in the
presence of some exogenous nucleating events.

We refer in particular to Ref.~\cite{SorOsorio10} for a short review of the main
results concerning the statistical properties of the space-time dynamics
of self-excited marked Hawkes conditional Poisson processes.

\subsection{Multivariate Hawkes processes}

The Multivariate Hawkes Process generalizes expressions (\ref{hyjuetgq2h56h2tgj}) into
the following general form for the conditional Poisson intensity for an event of type $j$ among a set
of $m$ possible types (see the document \cite{Liniger09} for an extensive review):
\begin{equation}
\lambda_j(t | H_t) =  \lambda_j^0(t) + \sum_{k =1}^m  \Lambda_{kj} \int_{(-\infty, t) \times {\cal R}}
h_j(t-s) ~g_k(x)  ~N_k(ds \times dx)~,
\label{hyjuetgq2ujuk42tgj}
\end{equation}
where $H_t$ denotes the whole past history, $ \lambda_j^0$ is
the rate of spontaneous (exogenous) events of type $j$, sources of immigrants of type $j$, $ \Lambda_{kj}$
is the $(k,j)$'s element of
the matrix of coupling between the different types which quantifies the ability of a type $k$-event
to trigger a type $j$-event. Specifically,
the value of an element $\Lambda_{jk}$ is just the average
number of first-generation events of type $j$ triggered by an event of type $k$.
The memory kernel $h_j(t-s)$ gives the probability that an event of type $k$ that
occurred at time $s<t$ will trigger an event of type $j$ at time $t$.
The function $h_j(t-s)$ is nothing but the distribution of waiting times (here between the
impulse of event $k$ which impacted the system at time $s$, the system taking
a certain time $t-s$ to react with an event of type $j$, this time  being a random
variable distributed according to the function $h_j(t-s)$.  The fertility (or productivity) law
$g_k(x)$ of events of type $k$ with mark $x$ quantifies the total average number of
first-generation events of any type triggered by an event of type $k$.
We have used the standard notation $ \int_{(-\infty, t) \times {\cal R}} f(t,x) N(ds \times dx) :=
\sum_{k | t_k < t} f(t_i, x_i)$.

The matrix ${\Lambda_{kj}}$ embodies both the topology of the network of interactions between
different types, and the coupling strength between
elements. In particular, ${\Lambda_{kj}}$  includes
the information contained on the adjacency matrix of the underlying network.
Analogous to the condition $n<1$ (subcritical regime) for the stability and stationarity of the monovariate Hawkes process,
the condition for the existence and stationarity of the process defined by (\ref{hyjuetgq2ujuk42tgj}) is that
the spectral radius of  the matrix ${\Lambda_{kj}}$ be less than $1$.
Recall that the spectral radius of a matrix is nothing but its largest eigenvalue.

\section{Multivariate generating function (GF)}

\subsection{Definition for events of first-generation events triggered by a given mother of type $k$}

Among the $m$ types of events, consider the $k$-th type and its first generation offsprings.
Let us denote $R_1^{k,1}, R_1^{k,2},\dots, R_1^{k,m}$, the number of ``daughter'' events
of first generation of type $1, 2, \dots, m$ generated by this ``mother'' event of type $k$.
With these notations, the generating function (GF) of all events of first generation
that are triggered by a mother event of type $k$ reads
\begin{equation}
A_1^k(y_1,y_2,\dots,y_m) := \text{E}\left[ \prod_{s=1}^m y_s^{R_1^{k,s}}\right] ~,
\end{equation}
where $\text{E}\left[ .\right]$ represents the statistical average operator.
One may rewrite this function in probabilistic form
\begin{equation}\label{1}
A_1^k(y_1,y_2,\dots,y_m) := \sum_{r_1=0}^\infty \dots \sum_{r_m=0}^\infty P_k(r_1,\dots,r_m) \prod_{s=1}^m y_s^{r_s}~,
\end{equation}
where $P_k(r_1,\dots,r_m)$ is the probability that the mother event of type $k$
generates $R^{k,1}=r_1$ first-generation events of type $1$, $R^{k,2}=r_2$ first-generation events of type $2$,
and so on. These probabilities satisfy to normalizing condition
\begin{equation}
\sum_{r_1=0}^\infty \dots \sum_{r_m=0}^\infty P_k(r_1,\dots,r_m) =1 ~.
\end{equation}

The first-order moments or mean values of the numbers of first-generation events of different types
triggered by a mother of type $k$ are given by
\begin{equation}\label{2}
n_{k,s} = {\partial\over\partial y_s} A_1^k(y_1,y_2,\dots,y_m)\big|_{y_1=\dots =y_m=1}
\end{equation}

\subsection{Generating function (GF) for all-generation events triggered by a given mother of type $k$}

The GF  $A^k(y_1,y_2,\dots,y_m)$  for all-generation events triggered by a given mother of type $k$
is by definition equal to
\begin{equation}
A^k(y_1,y_2,\dots,y_m) := \text{E}\left[ \prod_{s=1}^m y_s^{R^{k,s}}\right] ~,
\end{equation}
where $R^{k,1},R^{k,2},\dots,R^{k,m}$ are the numbers of events of all generations and all kinds
that are triggered by the mother event of type $k$.

In order to relate $A^k(y_1,y_2,\dots,y_m)$ to $A_1^k(y_1,y_2,\dots,y_m)$, we assume that
the first-generation daughters can also trigger their own daughters (which are the grand-daughters
of the initial event) according to the following rules.
\begin{itemize}
\item The numbers of second-generation events that are triggered by each first-generation event are statistically independent
of the numbers of first-generation events. They are also statistically independent of the numbers of second-generation events
that are triggered by any other first-generation events.
\item Each first-generation event triggers second-generation events according to the same laws controlling
the triggering of first-generation events by the initial mother event of the same type. In other words, the
same laws apply to the generation of new events from generation to generation, independently of the
generation depth.
\end{itemize}

These rules allow us to derive the GF of the numbers of first-generation and of second-generation events
by performing the following replacement for each variables $y_q$ in
the expression of the GF $A_1^k(y_1,y_2,\dots,y_m)$:
\begin{equation}
y_q \qquad \rightarrow \qquad y_q \cdot A_1^q(y_1,y_2,\dots,y_m)~ .
\end{equation}
The GF of the numbers of first-generation and of second-generation events that are triggered by
a mother event of type $k$ is given by the following expression in terms of the
GF $A_1^k(y_1,y_2,\dots,y_m)$ of the numbers of first-generation events that are triggered by
a mother event of type $k$:
\begin{equation}
A_2^k(y_1,y_2,\dots,y_m) =
A_1^k(y_1\cdot A_1^1(y_1,y_2,\dots,y_m),\dots,y_m\cdot A_1^m(y_1,y_2,\dots,y_m)) ~.
\end{equation}
This equation is valid for all possible values of $k = 1,\dots, m$.

By recurrence, one obtain the GF $A^k_{j+1}$ of the numbers of events
of all generations up to $j+1$ that are triggered by an initial mother event of type $k$
as a function of the GF $\{A^k_{j}\}$ of the numbers of events
of all generations up to $j$ triggered by an initial mother event of type $k$:
\begin{equation}
A_{j+1}^k(y_1,y_2,\dots,y_m) =
A_1^k(y_1\cdot A_j^1(y_1,y_2,\dots,y_m),\dots,y_m\cdot A_j^m(y_1,y_2,\dots,y_m))~.
\label{htyjyq}
\end{equation}
This equation is valid for all possible values of $k = 1,\dots, m$ and for all possible
generation levels $j=2$ to $+\infty$.

We assume that the above set of recurrence equations (\ref{htyjyq}) for $k = 1,\dots, m$
converges to some set of GF's $\{A^k(y_1,y_2,\dots,y_m); k = 1,\dots, m\}$. Then,
the corresponding GF's  $\{A^k(y_1,y_2,\dots,y_m); k = 1,\dots, m\}$
are solutions of the transcendent equations
\begin{equation}
A^k(y_1,y_2,\dots,y_m) =
A_1^k(y_1\cdot A^1(y_1,y_2,\dots,y_m),\dots,y_m\cdot A^m(y_1,y_2,\dots,y_m)) ~,~ k = 1,\dots, m~.
\label{jkik4w}
\end{equation}
The equation constitutes the basis for our subsequent analysis.

\section{Stability conditions using mean numbers of events of all generations}

\subsection{General relations}

The statistical average of the total numbers of events of type $s$ over
all generations that are triggered by a mother
of type $k$ is given by
\begin{equation}\label{meantotalrks}
\bar{R}^{k,s} = {\partial\over\partial y_s} A^k(y_1,y_2,\dots,y_m)\big|_{y_1=y_2 =\dots =y_m=1} ~.
\end{equation}
Using (\ref{jkik4w}), it is straightforward to show that $\bar{R}^{k,s}$ is solution of
\begin{equation}
\bar{R}^{k,s} = n_{k,s} + \sum_{\ell=1}^m n_{k,\ell}\cdot \bar{R}^{\ell,s}~,
\label{thjyjkk5m}
\end{equation}
where $n_{k,s}$ is the mean number of first-generation events of type $s$ triggered
by the mother event of type $k$.

Since expression (\ref{thjyjkk5m}) holds for all $k = 1,\dots, m$ and $s = 1,\dots, m$, it can
written in matrix form
\begin{equation}
\label{3}
 \hat{R} = \hat{N}+ \hat{N} \hat{R}~ ,
\end{equation}
where $\hat{N} = [n_{k,s}]$ is the matrix of the mean numbers of first-generation
events and $\hat{R} = [\bar{R}^{k,s}]$ is the matrix of the mean numbers of events
over all generations.
The sum over row indices of the elements of the matrix $\hat{N}$
\begin{equation}
n_k = \sum_{s=1}^m n_{k,s}~ ,
\label{tjetmnw}
\end{equation}
is the mean number of first-generation events of all kinds that are triggered by
a mother event of type $k$.

The solution of the matrix equation \eqref{3} is
\begin{equation}\label{rsol}
\hat{R} = {\hat{N}\over \hat{I}-\hat{N}}~ .
\end{equation}

The rest of the paper is concerned with the analysis of particular examples of this general solution,
worked out for different systems and excitation conditions embodied in
different forms of the matrix $\hat{N}$ of the mean numbers of first-generation
events.

\subsection{Symmetric mutual excitations}

Let us consider the case where
\begin{equation}
n_{k,k} = a ~; \qquad  \qquad n_{k,s} = b , ~~ k\neq s ~,
\label{kumnw}
\end{equation}
resulting in the form
\begin{equation}\label{nmatrixab}
\hat{N} = \left[
\begin{array}{c}
a ~~ b ~~ b ~~ ............... ~b
\\
b~~ a ~~ b ~~ ............... ~b
\\
b~~ b ~~ a ~~............... ~b
\\
.............................
\\
\underbrace{\mathstrut b.................~ b ~~ b ~~ a}_m
\end{array}
\right]
\end{equation}
for the matrix $\hat{N}$ of the mean numbers of first-generation  events.
This form (\ref{kumnw}) means that events of a given type have identical triggering efficiencies quantified by $a$
to generate first-generation events of the same type.  They also have identical efficiencies quantified by $b$
to trigger first-generation events of a different type. In other words,
the mean number of first-generation events of type $k$  triggered by a mother event
of the same type $k$ is independent of $k$. And the mean number of first-generation
events of any type $s \neq k$  triggered by any another event of a different type $k$ is independent of $k$ and $s$.

As a consequence, the mean number of first-generation events of all kinds that are triggered by
a mother event of some type $k$, as given by (\ref{tjetmnw}), is independent of $k$ and given by
\begin{equation}\label{ambn}
n_k  =  n = a + (m-1) b ~, ~~~~{\rm for~all}~k~.
\end{equation}

It is convenient to introduce the factor
\begin{equation}\label{qba}
q = {b\over a}
\end{equation}
comparing the inter-types with the intra-type triggering efficiencies.
Using definition \eqref{qba} and equality \eqref{ambn}, we obtain
\begin{equation}\label{abdefs}
a = {n\over 1+(m-1) q} ~, \qquad b = {n q \over 1+(m-1) q}~ .
\end{equation}
Two limiting cases are worth mentioning:
$q=0$ (independent types) and $q=1$ (fully equivalent types):
\begin{equation}
a|_{q=0} = n , \quad b|_{q=0} = 0 , \qquad a|_{q=1} = b|_{q=1}= {n\over m} ~.
\end{equation}

The solution (\ref{rsol}) implies that the matrix $\hat{R}$
possesses the same structure as the matrix $\hat{N}$, with identical diagonal elements $\bar{R}^{k,k}$
and identical off-diagonal elements $\bar{R}^{k,s}$ (for $k \neq s$), given respectively by
\begin{equation}\label{abforr}
\begin{array}{c}\displaystyle
\bar{R}^{k,k} ={n\over 1-n} \cdot {1+ n(q-1)\over 1+n(q-1) + q (m-1)} ~ ,  ~~ k=1, ..., m ~,
\\[4mm] \displaystyle
 \bar{R}^{k,s} = {n\over 1-n} \cdot {q\over 1+n(q-1) + q (m-1)} , \qquad k\neq s.
\end{array}
\end{equation}
Therefore, the mean $\bar{R}^k $ of the total number of events of all kinds, that are triggered by some given mother event
of a given type $k$, is given by
\begin{equation}\label{rkssum}
\bar{R}^k =  \sum_{s=1}^m \bar{R}^{k,s}  = {n\over 1-n} := \bar{R}~, ~~~\forall q~.
\end{equation}
This expression has a simple interpretation, resulting from equation (\ref{ambn}) and the process
of triggering. Indeed, by definition of $n$ in (\ref{ambn}), there are on average $n$
first-generation events of all kinds that are triggered by
a mother event of some type $k$. Each of these first-generation event triggers on average $n$
second-generation events of all kinds, leading to a total contribution $n^2$ for the number
of second-generation events. Counting all the generation cascades, we obtain
$n + n^2 + n^3 + ...$, which is nothing but the result (\ref{rkssum}).

As for the mono-variate Hawkes process, the dynamics is stable (sub-critical) for $n<1$
and unstable (super-critical or exponentially explosive) for $n>1$. As usual, the critical point
occurs when there is exactly $n=1$ first-generation events of all kinds that are triggered by
a mother event of any type. There is not qualitative difference between this
multi-variate Hawkes process with the structure (\ref{nmatrixab}) of mutual excitations
and a mono-variate Hawkes process, once the branching ratio $n$ defined as
the average number of first-generation daughters from a given mother is generalized into
its natural extension (\ref{ambn}).

\subsection{Two-dimensional mutually and self-excited Hawkes process \label{jrtwbqq}}

With only two types of events, a detailed analysis can be performed, with
the discovery of new qualitative regimes.

\subsubsection{Stability analysis}

With two types of events, the $2 \times 2$ matrix $\hat{N}$ of the mean numbers of of first-generation events
can be kept fully general and is noted as
\begin{equation}\label{twodimnmatr}
\hat{N} = \left[
\begin{array}{c}
n_{1,1} ~ ~ n_{1,2}
\\[2mm]
n_{2,1} ~ ~ n_{2,2}
\end{array} \right]
\end{equation}
The solution (\ref{rsol}) is a $2 \times 2$ matrix $\hat{R}$ with elements $\bar{R}^{k,s}$ given by
\begin{equation}\label{rkstwodim}
\begin{array}{c} \displaystyle
\bar{R}^{1,1} = {n_{1,1} + n_{1,2} n_{2,1} - n_{1,1} n_{2,2} \over \mathcal{D}} , \qquad \bar{R}^{1,2} = {n_{1,2} \over \mathcal{D}} ,
\\[4mm]\displaystyle
\bar{R}^{2,1} = {n_{2,1} \over \mathcal{D}} , \qquad
\bar{R}^{2,2} = {n_{2,2} + n_{1,2} n_{2,1} - n_{1,1} n_{2,2} \over \mathcal{D}} ,
\end{array}
\end{equation}
where
\begin{equation}\label{mathddef}
\mathcal{D} = 1+n_{1,1} n_{2,2} - n_{1,2} n_{2,1} - n_{1,1} - n_{2,2} .
\end{equation}

In order to determine the stability of two-dimensional mutually and self-excited Hawkes process, we study
the mean numbers $n_1$ and $n_2$ of first-generation
events triggered by a mother event of the first and second kind, respectively. They are given
by the sums of the two row elements of the matrix $\bar{N}$:
\begin{equation}
 n_1 = n_{1,1} + n_{1,2}  , \qquad   n_2 = n_{2,1} + n_{2,2}  .
\end{equation}
It is convenient to use a representation of  the elements $n_{k,s}$ of the matrix $\hat{N}$ similar to
\eqref{abdefs}:
\begin{equation}\label{nonetwoqdef}
n_{1,1} = {n_1\over 1+q_1}~ , \quad n_{1,2} = {n_1 q_1\over 1+q_1} ~, \quad n_{2,1} = {n_2 q_2\over 1+q_2} ~, \quad n_{2,2} = {n_2\over 1+q_2} ~.
\end{equation}
As in (\ref{qba}), $q_1$ and $q_2$ quantify the
relative strengths of inter-type compared with the intra-type triggering efficiencies:
$q_1 = n_{1,2}/n_{1,1}$ and $q_2 = n_{2,1}/n_{2,2}$. The limit $q_1=q_2=0$ reduces
to two independent self-excited Hawkes processes.

The solutions $\bar{R}^{k,s}$ given by (\ref{rkstwodim}) are finite as long as
the spectral radius $\lambda(n_1, n_2)$ of the matrix $\bar{N}$ remains smaller than $1$.
This defines the sub-critical regime. The system becomes critical (respectively super-critical)
when the spectral radius $\lambda(n_1, n_2)$ is equal to $1$ (respectively larger than $1$).
One can show that the set $(n_1; n_2)$ such that the denominator
$\mathcal{D}$ given by (\ref {mathddef}) is identically zero is critical, i.e., corresponds to
a unit spectral radius, if $\mathcal{D}$ remains positive in the domain
bounded by the semi-axes $[n_1 \in (0;+\infty), n_2 \in (0;+\infty)]$
and the curve $\mathcal{D}(n_1; n_2) = 0$.

In order to study the three regimes (sub-critical, crtical and super-critical),
it is convenient to express $\mathcal{D}(n_1,n_2)$ as a function of $n_1$ and $n_2$,
using (\ref{nonetwoqdef}):
\begin{equation}\label{mathdqexpr}
\mathcal{D}(n_1,n_2) = 1+ {n_1 n_2 (1-q_1 q_2)\over (1+q_1)(1+q_2)} - {n_1\over 1+q_1} - {n_2\over 1+q_2} ~.
\end{equation}

\begin{quote}
\centerline{
\includegraphics[width=11cm]{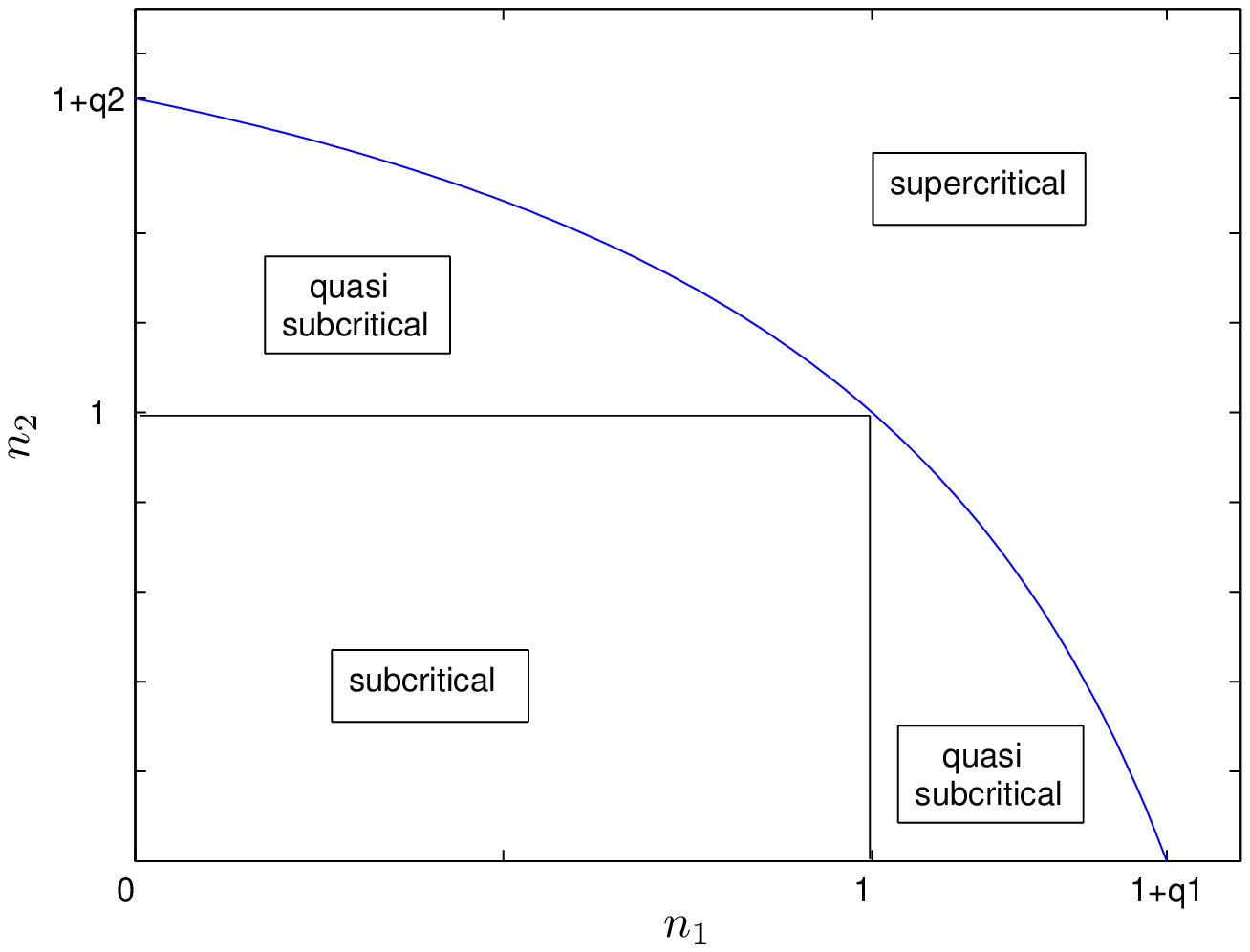}}
{\bf Fig.~1:} \small{Critical line $\mathcal{D}(n_1,n_2) =0$ and three domains in the plane $(n_1,n_2)$: 1) subcritical, where both $n_1$ and $n_2$ are smaller than $1$; 2) quasi subcritical, where one of the mean numbers $(n_1,n_2)$
of first-generation events is larger than one but the mean numbers of events of all generations are finite; 3) supercritical region, where all mean numbers $\bar{R}^{k,s}$ are infinite.}
\end{quote}

The critical line $\mathcal{D}(n_1,n_2) =0$ is shown in figure~1 in the plane $(n_1, n_2)$, together with three
domains.
\begin{enumerate}
\item For $n_1<1$ and $n_2<1$, $\mathcal{D}(n_1,n_2) >0$ and the system is subcritical. The
conditions $n_1<1$ and $n_2<1$ mean that the cascade of events over all generations do not blow up
for each of the two types of event triggering.

\item The domain indicated in figure~1 as ``quasi subcritical'' is such that one of the mean numbers $(n_1,n_2)$
of first-generation events is larger than one but the mean numbers of events over all generations remain finite
since $\mathcal{D}(n_1,n_2) > 0$. Intuitively, the supercritical regime of one of the event types is damped
out by the triggering of the second type of events which is subcritical. The two extreme boundaries
$(n_1=1+q_1; n_2=0)$ and $(n_1=0; n_2=1+q_2)$ exemplify this point as they correspond respectively to $n_{1,1}=1; n_2=0$
and $n_1=0; n_{2,2}=1$.

\item In the domain indicated at ``supercritical'' in figure~1, the mean number of
events of both types summed over all generations goes to infinity. This
occurs of course if both $n_1$ and $n_2$ are larger than $1$ but also when one of them is smaller than $1$
if the other one is sufficiently large. In this later case, the damping offered by the second type is not sufficient
to stabilize the triggering process. This is the runaway explosive regime.

\item The downward sloping line defines the critical domain $\mathcal{D}(n_1,n_2) =0$ separating
the quasi subcritical and the supercritical regimes.

\end{enumerate}

As the critical line $\mathcal{D}(n_1,n_2) =0$ is approached from within the subcritical regime,
the total mean number $\bar{R}^1 = \bar{R}^{1,1}+\bar{R}^{1,1}$ of events of all types and over all generations,
\begin{equation}
\bar{R}^1 = \bar{R}^{1,1}+\bar{R}^{1,1} =
{n_1(1+q_1+q_2 +q_1 q_2 -n_2 (1-q_1 q_2))\over (1+q_2) (1+q_1-n_1)-(1+q_1-n_1 (1-q_1 q_2)) n_2}~,
\label{thyju7ki7k}
\end{equation}
grows to finally diverge on the line, as shown in figure~2 for a particular example.

In this example, $q_1=0.2, q_2 =0.4, n_1=0.8$. The value $q_1/(1+q_1) = 16.7\%$ is the fraction
of first-generation events generated by a mother of the first type which are of the second type.
The value $q_2/(1+q_2) = 28.6\%$ is the fraction
of first-generation events generated by a mother of the second type which are of the first type.

\begin{quote}
\centerline{
\includegraphics[width=11cm]{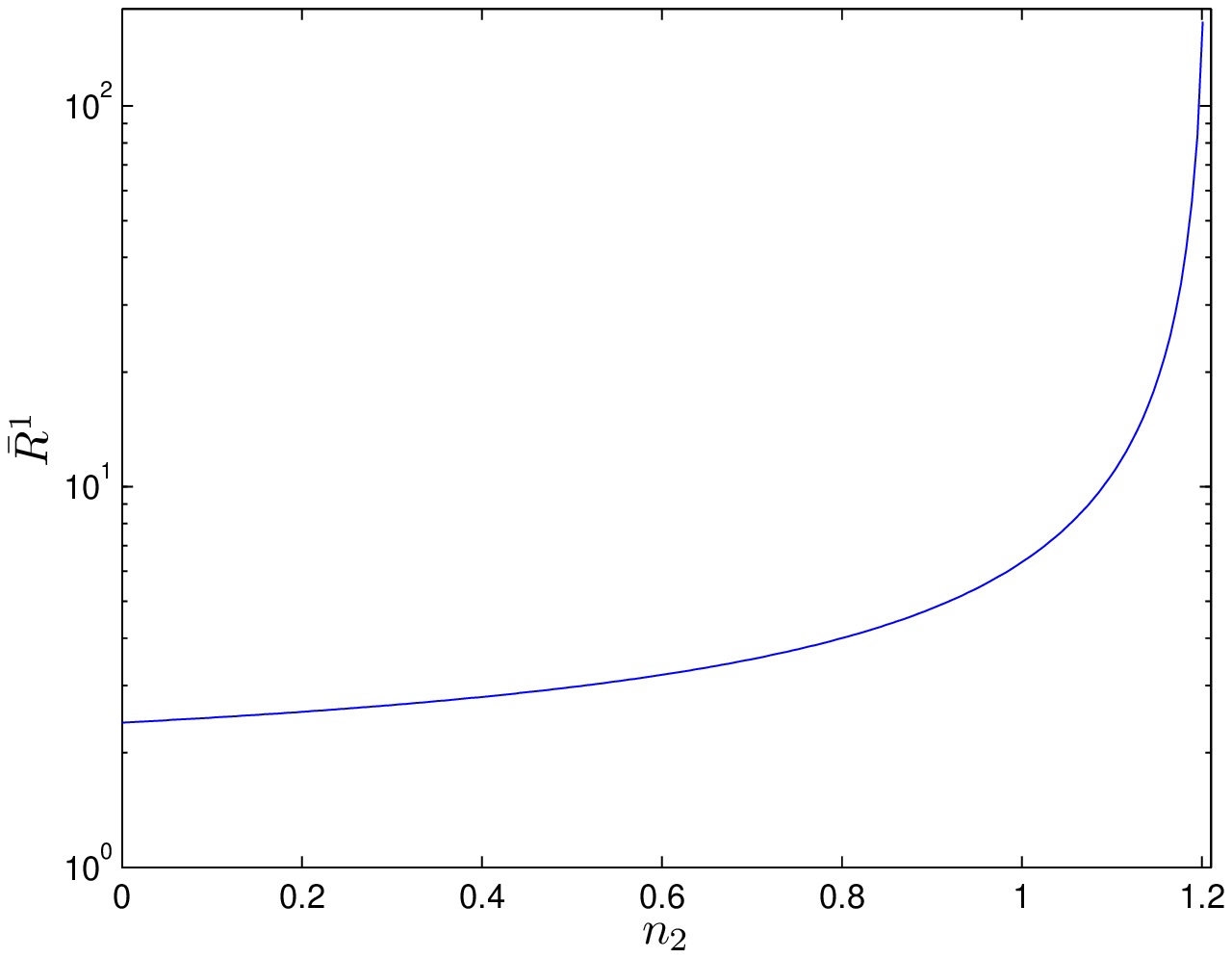}}
{\bf Fig.~2:} \small{Linear-log plot of the mean number $\bar{R}^1$ of events of all types and over all generations
given by expression (\ref{thyju7ki7k}) as a function of $n_2$ for
$q_1=0.2$, $q_2=0.4$ and $n_1=0.8$. Note that $\bar{R}^1$ remains finite even when $n_2$ becomes larger than $1$
up to a critical value $n_2^c = 1.206896...$ for which $\mathcal{D}(n_1,n_2) =0$ at which it diverges. For $n_2=1.2$
for instance, $\bar{R}^1=144$. }
\end{quote}

\subsubsection{Strong asymmetry in mutual triggering \label{hyjuj4uj}}

It is instructive to consider the limiting case where one type of events triggers many more events
of the other type than the reverse. Mathematically, this corresponds to
\begin{equation}
q_1\gg q_2 ~ ,
\end{equation}
which means that the fraction of first-generation events generated by a mother of the first type which are of the second type
is much larger than the fraction of first-generation events generated by a mother of the second type which are of the first type.
In this case, the critical line $\mathcal{D}(n_1,n_2)=0$ becomes almost rectangular, as illustrated in figure~3.

\begin{quote}
\centerline{
\includegraphics[width=11cm]{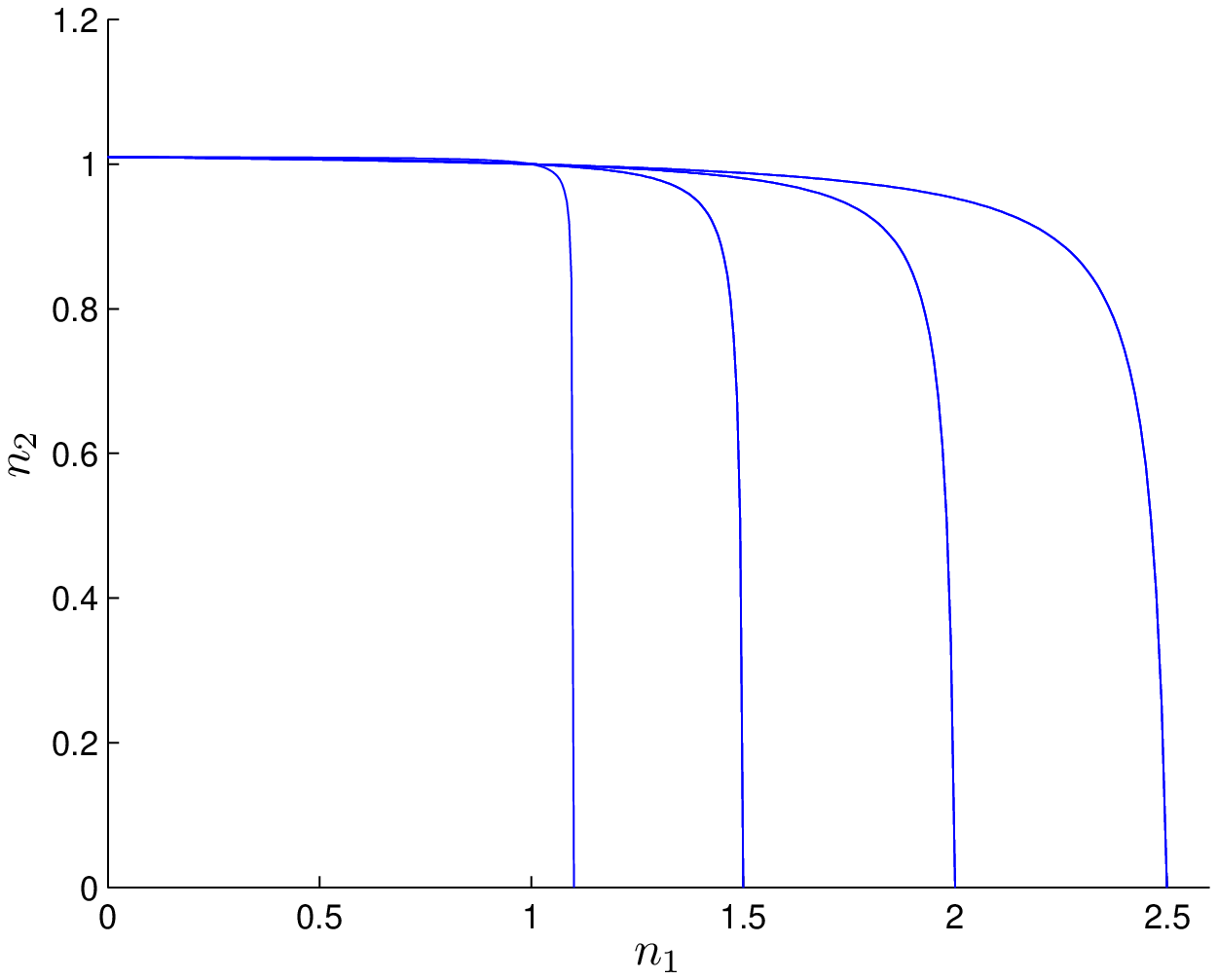}}
{\bf Fig.~3:} \small{Plots of the critical line $\mathcal{D}(n_1,n_2)=0$ for the four following
values of $(q_1;q_2)$: (0.1;0.01), (0.5;0.01), (1;0.01), (1.5;0.01) from left to right.}
\end{quote}

Let us consider in more details the limiting case $q_1=q$  while $q_2=0$, for which
the relations \eqref{nonetwoqdef} transform into
\begin{equation}
\label{nonetwoqzerodef}
n_{1,1} = {n_1\over 1+q}~ , \quad n_{1,2} = {n_1 q\over 1+q} ~, \quad n_{2,1} = 0~ , \quad n_{2,2} = n_2~ .
\end{equation}
As a result, the relations \eqref{rkstwodim} and \eqref{mathddef} become
\begin{equation}\label{rksqtwozero}
\begin{array}{c}\displaystyle
\bar{R}^{1,1} = {n_1\over 1+q -n_1} , \qquad \bar{R}^{1,2} = {n_1 q\over (1+q-n_1) (1-n_2)} ,
\\[4mm]\displaystyle
\bar{R}^{2,2} = {n_2\over 1-n_2} , \qquad \bar{R}^{2,1} = 0 .
\end{array}
\end{equation}

The finiteness (subcritical behavior) of the number $\bar{R}^{1,1}$ is controlled solely by
$n_1$, which must be smaller than $1+q$. As $n_1$ tends to $1+q$,  $\bar{R}^{1,1}$ goes
to infinity, expressing the transition to the supercritical regime.
In contrast, there are two mechanisms leading to the divergence of $\bar{R}^{1,2}$.
\begin{enumerate}
\item As $n_1$ tends to $1+q$, $\bar{R}^{1,2}$ goes to infinity, as the number of
events of the first type itself diverges, each of these events triggering a significant fraction
of events of the second type, at each generation. In other words, the divergence of $\bar{R}^{1,2}$
is controlled by or slaved to that of $\bar{R}^{1,1}$, which reflects the triggering efficiency
of events of type one.

\item The number $\bar{R}^{1,2}$ of events of the second type generated over all
generations by a mother event of type one diverges when $n_2 \to 1$, even
if $n_1 < 1+q$.   A mother event of type one triggers events of type two at each generation, each of these events
only triggering events of their own kind. Thus, the number of events of the second type
diverges when the self-triggering parameter (or ``branching ratio'') $n_2$ reaches its critical value $n_2^c=1$.
\end{enumerate}
This implies in particular that, when $q$ or $n_2$ are sufficiently large, the following
inequality holds: $\bar{R}^{1,2} > \bar{R}^{1,1}$.  The general condition for this to be true is  $n_2+q>1$.
Intuitively, for a fixed self-triggering ability $n_2$, there must be sufficiently many events of type two
generated by events of type one: $q > 1-n_2$. Alternatively, for a fixed fraction $q/(1+q)$ of
first-generation events of type two generated by events of type one, the branching ratio of type two events
must be sufficient large: $n_2 > 1-q$.

Note that, if both $n_{12}$ and $n_{21}$ are positive, corresponding to a non-degenerate case,
then events of any type do trigger events of the other type. As a consequence,
either $\bar{R}^{1}$ and $\bar{R}^{2}$ are both finite or both infinite. This is not the case
for independent or semi-independent systems. A system is independent
if $n_{12} = n_{21} = 0$, i.e., events of different types live their separate ``lives'' without
any inter-mutual triggering. In this case, for
instance, if $n_1 < 1$, while $n_2 > 1$, then the events of the first type form
a subcritical set, while the events of the second type form
a supercritical system. Two systems are semi-independent if only one
of the two cross-terms $n_{12}$ and $n_{21}$ is equal to zero. Suppose
for instance that $n_{21} = 0$. Then, if $\bar{R}^{1,1}$ is finite, then
$\bar{R}^{1,2}$ might be finite or infinite,
just because events of the second type cannot trigger events of the second type
and an infinite value of $\bar{R}^{1,2}$ remains compatible with a finite value of $\bar{R}^{1,1}$.
In contrast, if $\bar{R}^{1,1}$ is infinite, then $\bar{R}^{1,2}$ is necessarily infinite,
because the infinite number of events of the first type trigger
an infinite number of events of the second type, since $n_{12} \neq 0$,
even if self-triggering of events of type two is zero ($n_{22} = 0$).
Moreover, a main peculiarity of degenerate (independent or semi-independent)
systems is a strongly rectangular critical curve.

\begin{quote}
\centerline{
\includegraphics[width=11cm]{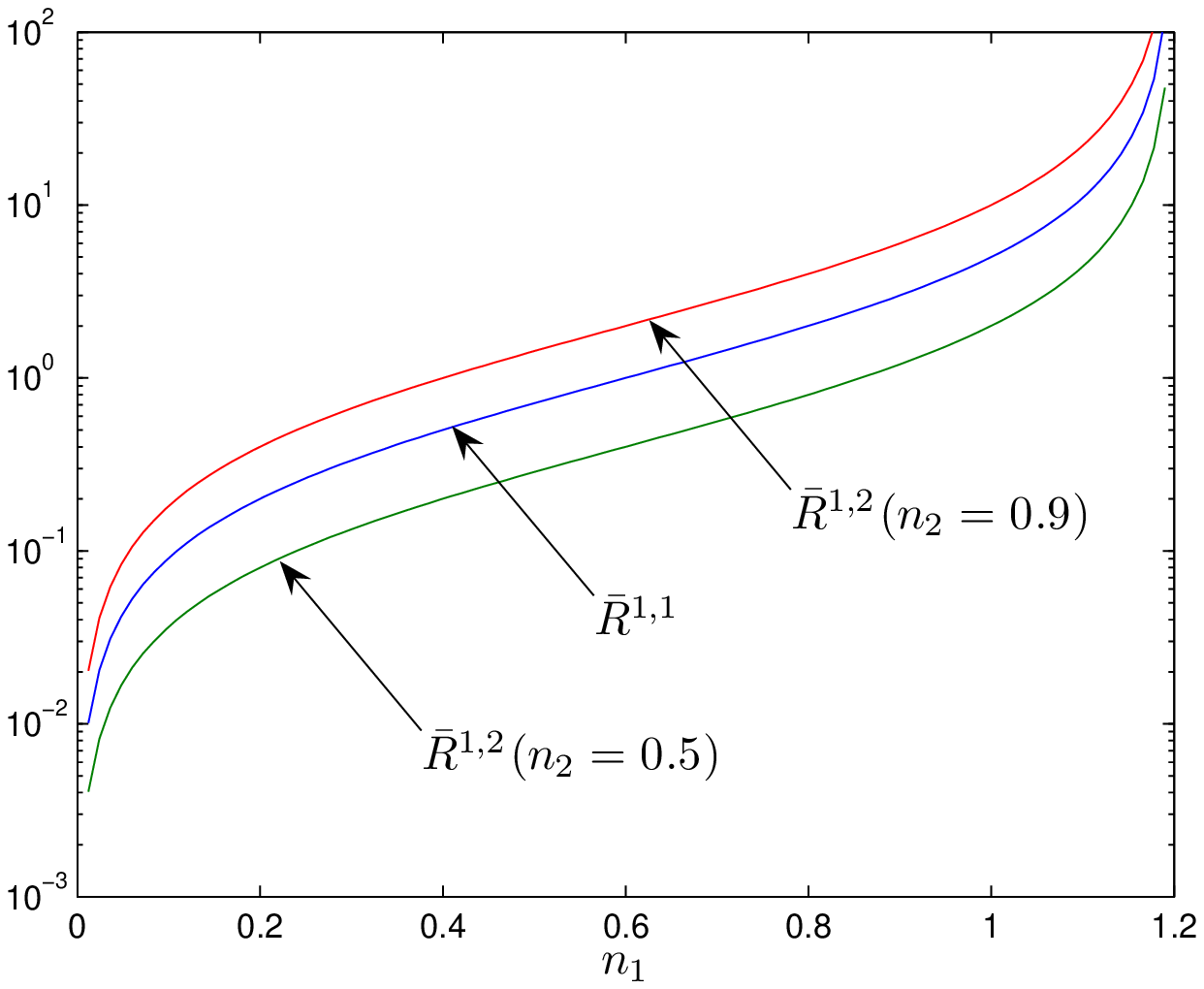}}
{\bf Fig.~4:} \small{(color online) Linear-log dependence of the mean numbers $\bar{R}^{1,1}$ and $\bar{R}^{1,2}$
given by \eqref{rksqtwozero} as a function of $n_1$, for $q=0.2$ and for two values $n_2=0.5$ and $n_2=0.9$}
\end{quote}

Figure~4 shows in linear-log scale the number $\bar{R}^{1,1}$ (respectively $\bar{R}^{1,2}$)  of events of the first type
(respectively second type) over all generations that are triggered by a mother event of the first type, given by \eqref{rksqtwozero},
as functions of $n_1$, for $q=0.2$ and for two values $n_2=0.5$ and $n_2= 0.9$.
One can observe that, for $n_2=0.9$, $\bar{R}^{1,2}$ is larger than  $\bar{R}^{1,1}$ for any $n_1\in(0,1.2)$.

\subsection{One-dimensional chain of directed triggering \label{yjurkryumre}}

The case of a strong asymmetry in mutual triggering discussed in subsection \ref{hyjuj4uj}
for the two-dimensional case can be generalized to the case of $m>2$ different event types.
We consider a chain of directed influences $k \to k+1$ where the events of type $k$
trigger events of both types $k$ and $k+1$ only, and this for $k=1, 2, ...,m$.
This is captured by a form of the matrix $\hat{N}$ which has only the diagonal and
the line above the diagonal with non-zero elements.

As the simplest example, consider first the matrix $\hat{N}$
\begin{equation}
\label{nmatrixablent}
\hat{N} = \left[
\begin{array}{c}
\chi ~~ s ~~ 0 ~~ 0 ~~ 0 ......... ~0 ...
\\
0 ~~ \chi ~~ s ~~ 0 ~~ 0......... ~0 ...
\\
0~~ 0 ~~ \chi ~~ s ~~ 0......... ~0 ...
\\
0~~ 0 ~~ 0 ~~ \chi ~~ s ......... ~0 ...
\\
..................................
\end{array}
\right]
\end{equation}
where
\begin{equation}\label{ablent}
\chi={n\over 1+q} ~, \qquad s = {nq\over 1+q} ~.
\end{equation}
Reduced to a two-dimensional system $m=2$, this corresponds
to the particular case of subsection \ref{hyjuj4uj} for which
$n_{2,2} = n_2 = n_1/(1+q)$ in the notations of expressions (\ref{nonetwoqzerodef}),
because the diagonal elements are taken all equal.
For $m=3$, a financial example is that the events of the first type
correspond to fundamental news, the events of the second type are the price jumps of a leading
market such as the US (assuming no feedbacks of prices on news)
and the events of the third type correspond to the price jumps of a secondary
market, such as the Russian stock market (assuming to effect of the Russian
market on the US market).

Assuming that the mother event is of type $k$,
the solution of equation (\ref{rsol}) for this case
 \eqref{nmatrixablent} with \eqref{ablent} is given by $\bar{R}^{s,s} = \bar{R}^{k,s} = \bar{R}^{s,k}=0$
 for $1 \leq s <k$ and
\begin{equation}\label{rkslent}
\begin{array}{c} \displaystyle
\bar{R}^{k,k} = {\chi\over 1-\chi} = {n\over 1+q-n} ,
\\[5mm] \displaystyle
\bar{R}^{k,s} = {s^{s- k}\over(1-\chi)^{s- k+1}} = (1+q) {(nq)^{s- k}\over (1+q-n)^{s- k+1}} , \qquad  s > k ~ .
\end{array}
\end{equation}
The fact that the critical point is solely controlled by a single critical value $n_c = 1/(1+q)$ results
from the fact that all diagonal elements of the matrix (\ref{nmatrixablent}) are equal.

\subsection{One-dimensional chain of nearest-neighbor triggering}

A natural extension to the above one-dimensional chain of directed triggering
discussed in section \ref{yjurkryumre} is to include the possibility of feedbacks from events
of type $k+1$ to type $k$. The simple example is to consider symmetry mutual excitations
confined to nearest neighbors in the sense of event types: $k \leftrightarrow k+1$.
Mathematically, this is described by a symmetric matrix $\hat{N}$ of the
average numbers  $n_{k,s}$ of first-generation events of different types
triggered by a mother of a fixed type. Figure~5 provides the geometrical sense of
matrix $\hat{N}$ \eqref{nmatrixchscircle} for $m=6$, where the circles represent the
six types of events and the arrows denote their mutual excitation influences.

\begin{quote}
\centerline{
\includegraphics[width=11cm]{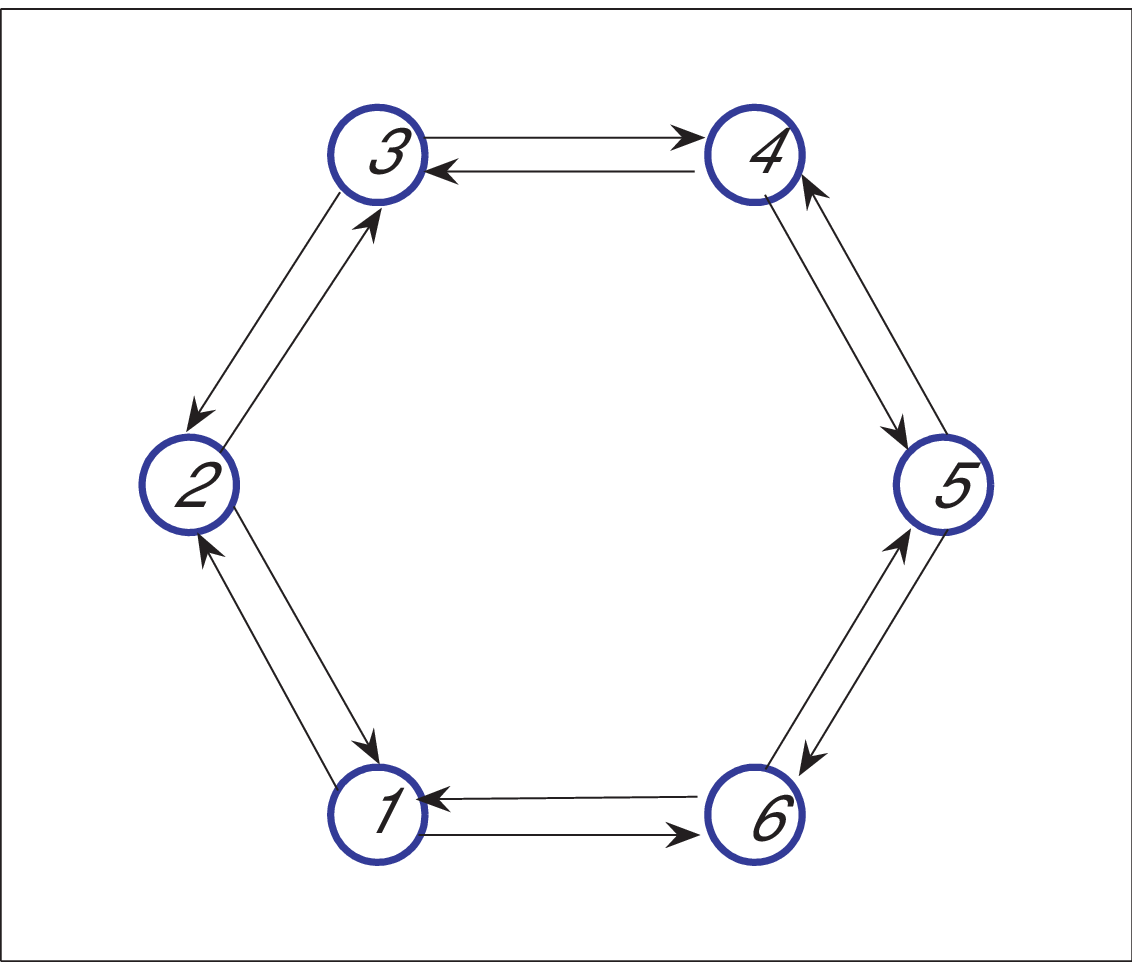}}
{\bf Fig.~5:} \small{Geometric sense of the matrix $\hat{N}$ for a one-dimensional chain of nearest-neighbor triggering.}
\end{quote}

Here, we consider that all diagonal elements are equal to some constant $\chi$ (same self-triggering abilities)
and all off-diagnoal elements are equal to some different constant $s$ (same mutual triggering abilities).
The elements $n_{1,m}$ and $n_{m,1}$ are also equal to $s$ to close the chain of mutual excitations
between events of type $1$ and of type $m$.
Restricting to $m=6$ for illustration purpose, the corresponding matrix $\hat{N}$ reads
\begin{equation}\label{nmatrixchscircle}
\hat{N} = \left[
\begin{array}{c}
\chi ~~ s ~~ 0 ~~ 0 ~~ 0 ~~ s
\\
s ~~ \chi ~~ s ~~ 0 ~~ 0 ~~ 0
\\
0 ~~ s ~~ \chi ~~ s ~~ 0 ~~ 0
\\
0 ~~ 0 ~~ s ~~ \chi ~~ s ~~ 0
\\
0 ~~ 0 ~~ 0 ~~ s ~~ \chi ~~ s
\\
s ~~ 0 ~~ 0 ~~ 0 ~~ s ~~ \chi
\end{array}
\right]
\end{equation}
where
\begin{equation}\label{chccirc}
\chi={n\over 1+q}~ , \qquad s =  {nq\over 2 (1+q)} \quad \Rightarrow \quad \chi+ 2s = n~.
\end{equation}
As before, the parameter $q$ quantifies the ``strength'' of the interactions
between events of different types. Here, $n$ represents the total number of first-generation
events of all types that are generated by a given mother of fixed arbitrary type.

The solution of equation (\ref{rsol}) for this case is given by a circulant structure
\begin{equation}\label{nmatritotalcircle}
\hat{R} = \left[
\begin{array}{c}
A ~~ B ~~ C ~~ D ~~ C ~~ B
\\
B ~~ A ~~ B ~~ C ~~ D ~~ C
\\
C ~~ B ~~ A ~~ B ~~ C ~~ D
\\
D ~~ C ~~ B ~~ A ~~ B ~~ C
\\
C ~~ D ~~ C ~~ B ~~ A ~~ B
\\
B ~~ C ~~ D ~~ C ~~ B ~~ A
\end{array}
\right]
\end{equation}
where
\begin{equation}\label{abcddef}
\begin{array}{c} \displaystyle
A(n,q) = \frac{4 n(1-n+q)^3-(1-n+q) (5n-2 q -2) n^2 q^2 - n^4 q^4}{ 4(1-n+q)^4- 5 (1-n+q)^2 n^2 q^2 +n^4 q^4} ~,
\\[6mm] \displaystyle
B(n,q) = \frac{n q (1+q) (2 (1-n+q)^2  - n^2 q^2)}{ 4(1-n+q)^4- 5 (1-n+q)^2 n^2 q^2 +n^4 q^4} ~,
\\[6mm] \displaystyle
C(n,q) = \frac{n^2 q^2 (1+q)(1-n+q)}{ 4(1-n+q)^4- 5 (1-n+q)^2 n^2 q^2 +n^4 q^4} ~,
\\[6mm] \displaystyle
D(n,q) = \frac{n^3 q^3 (1+q)}{ 4(1-n+q)^4- 5 (1-n+q)^2 n^2 q^2 +n^4 q^4} ~ .
\end{array}
\end{equation}

It is straightforward to check that the common denominator $4(1-n+q)^4- 5 (1-n+q)^2 n^2 q^2 +n^4 q^4$
to these four numbers $A(n,q), B(n,q), C(n,q)$ and $D(n,q)$
\begin{enumerate}
\item does not vanish for any $q$ values  for $n<1$,
\item vanishes for any $q$ values at $n=1$, and
\item can vanish at up to four values of $q$ for $n>1$.
\end{enumerate}
This means that the system resulting from the structure of mutual excitations between different event types represented by
the matrix (\ref{nmatritotalcircle}) is always in the subcritical regime for $n<1$
and becomes critical at $n=1$ as for the decoupled or monovariate case.

The behavior of the ratios
\begin{equation}\label{ratiosdef}
\mathcal{R}_2 = \frac{\bar{R}^{1,2}}{\bar{R}^{1,1}} = \frac{B}{A} , \quad \mathcal{R}_3 = \frac{\bar{R}^{1,3}}{\bar{R}^{1,1}} = \frac{C}{A} , \quad \mathcal{R}_4 = \frac{\bar{R}^{1,4}}{\bar{R}^{1,1}} = \frac{D}{A}
\end{equation}
is shown in figure~6 as a function of $n$ for $q=0.1$. Here, the mother event is of type 1 and the curves illustrate
the progressing dampening of the cascade of triggering proceeding from type to type via nearest-neighbor mutual
excitations.

\begin{quote}
\centerline{
\includegraphics[width=11cm]{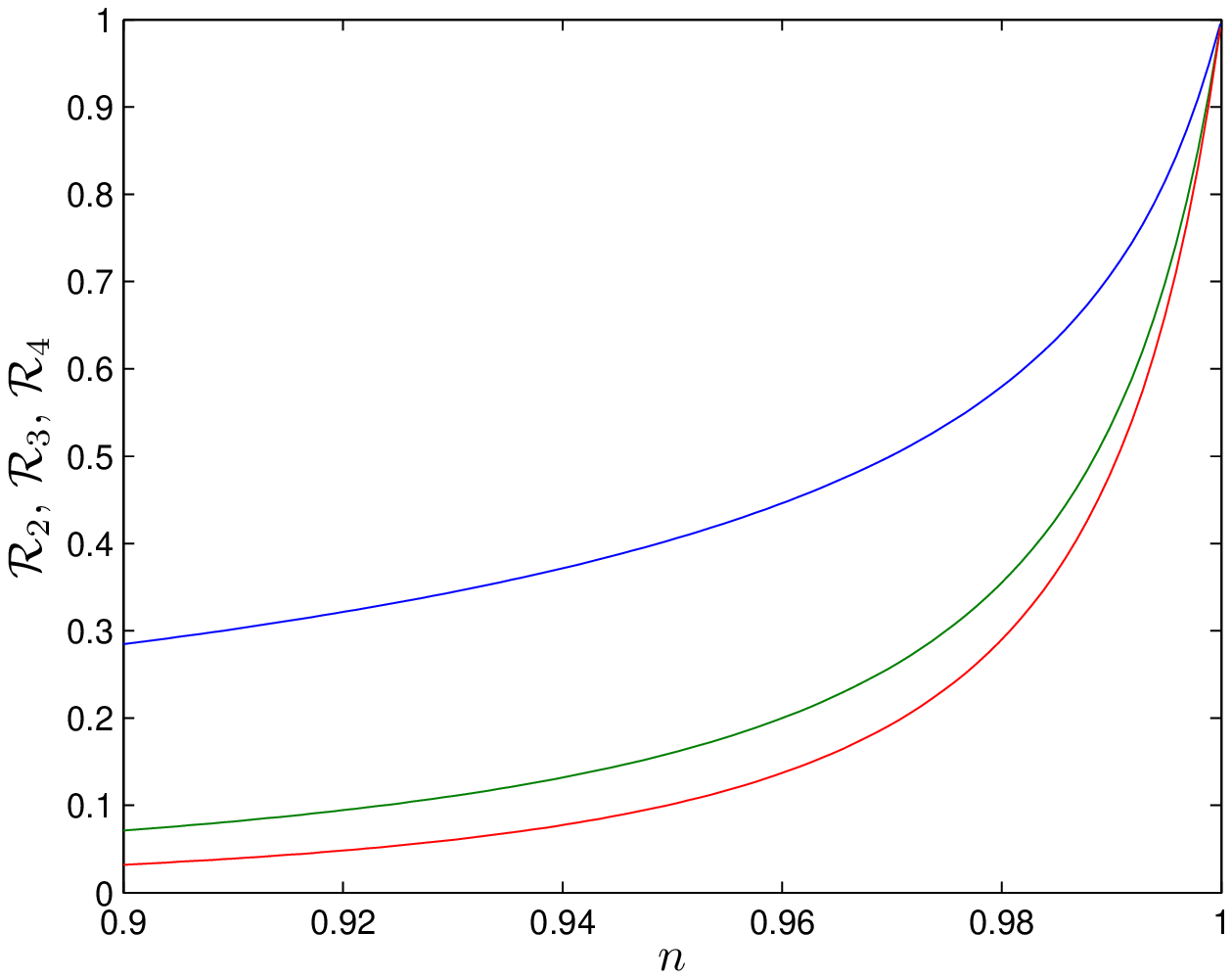}}
{\bf Fig.~6:} \small{(color online) Top to bottom: ratios $\mathcal{R}_2$, $\mathcal{R}_3$ and $\mathcal{R}_4$ defined in \eqref{ratiosdef} of
the mean numbers of events of all generations generated by a mother event of type $1$ for $q=0.1$.}
\end{quote}

\subsection{Subcriticality measure in the case of two types of events}

For a system with two types of events, the explicit relation defining the critical curve is
\begin{equation}
n_2 = \mathcal{G}(n_1,q_1,q_2)=
\frac{(1+q_1) (1+q_2) - (1+q_2) n_1}{(1+q_1)-(1-q_1 q_2) n_1} , \qquad 0\leqslant n_1 \leqslant 1+q_1 ,~
\end{equation}
which is represented in figure 7. One can verify that the point $(n_1=1,n_2=1)$ is always on
the critical curves.

\begin{quote}
\centerline{
\includegraphics[width=11cm]{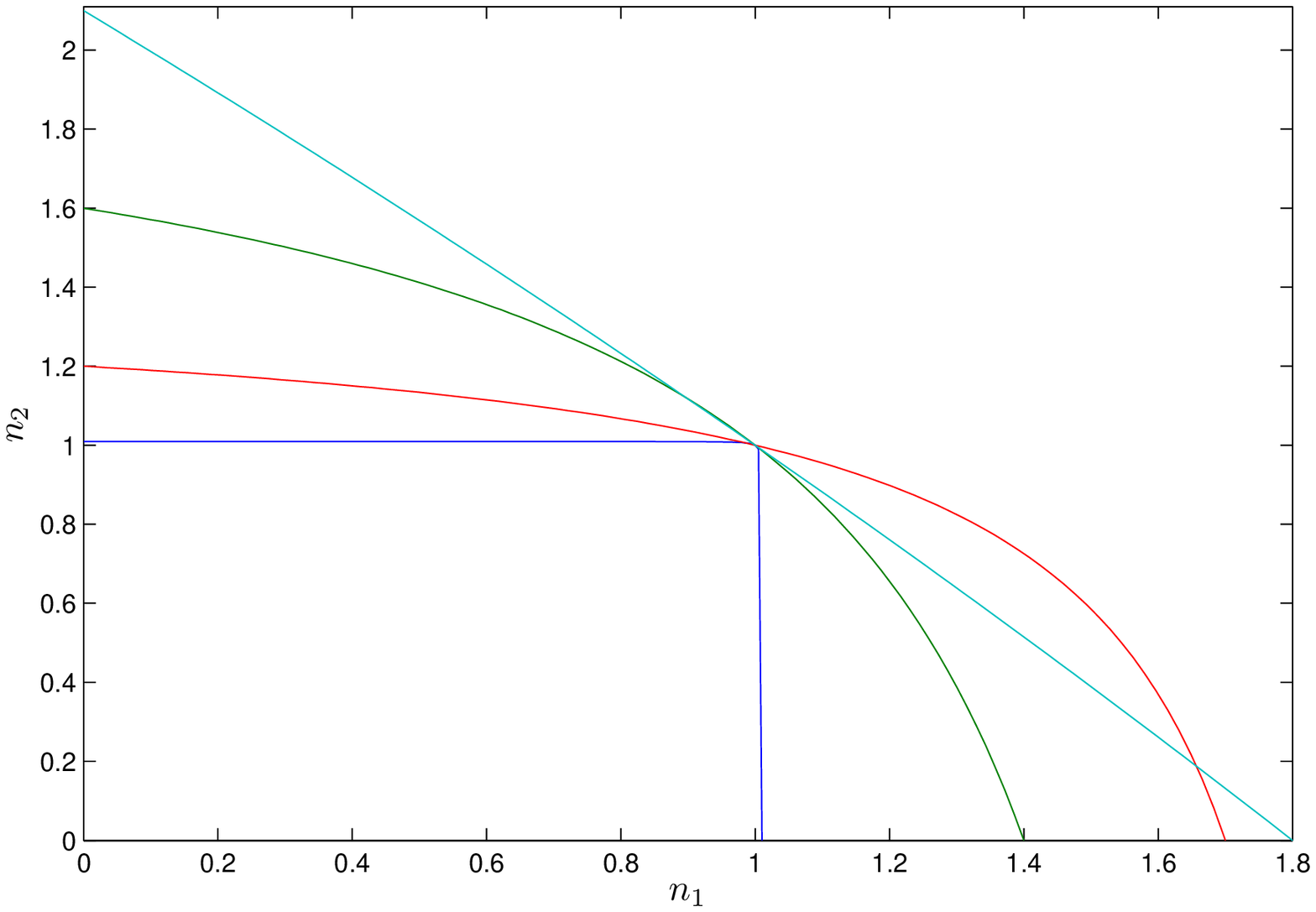}}
{\bf Fig.~7:} \small{(color online) Plots of critical curves $n_2(n_1)$ for systems with two types of events,
for the following pairs of parameters: bottom up on the left side of the curves, we have  ($q_1=0, q_2=0$), ($q_1=0.4, q_2=0.6$), ($q_1=0.7, q_2=0.2$) and for ($q_1=0.8, q_2=1.1$).}
\end{quote}

One can observe that,
for $q_1>0$ and $q_2>0$, the subcritical domain is significantly larger than for systems
in which event types are independent, i.e., do not mutually trigger each other, corresponding
to $q_1=q_2\equiv 0$. It is illuminating to introduce a quantitative measure of
the domain of subcriticality, here chosen for the two-dimensional case
as the surface $\mathcal{S}$ of the subcritical domain:
\begin{equation}
\mathcal{S}(q_1,q_2) = \int_0^{1+q} \mathcal{G}(n_1,q_1,q_2) d n_1~ .
\end{equation}
For independent types ($q_1=q_2\equiv 0$), $\mathcal{S}(0,0)=1$.
The  general expression of the subcriticality measure $\mathcal{S}(q_1,q_2)$ is obtained as
\begin{equation}
\mathcal{S}(q_1,q_2) = (1+q_1) (1+q_2) \frac{1- q_1 q_2 \left[1-\ln(q_1 q_2)\right]}{(1-q_1 q_2)^2}~ .
\end{equation}
In the particular case of a chain of directed triggering in the space of event types
studied in subsection \ref{yjurkryumre} for which the critical curve is rectangular, we find
\begin{equation}
\mathcal{S}(q):= \lim_{q_2\to 0_+} \mathcal{S}(q,q_2) = 1 + q~ .
\label{yjuj}
\end{equation}
In the case of symmetric triggering among different event types $(q_1=q_2=q)$, we obtain
\begin{equation}
\mathcal{S}(q,q) = \frac{1-q^2 + q^2 \ln(q^2)}{(1-q)^2}~ .
\label{tu44j}
\end{equation}
These last two functions $\mathcal{S}(q)$ and $\mathcal{S}(q,q)$ are depicted
in figure 8. The main insight is that mutual triggering tends to provide
a significant extension of the stability domain.

\begin{quote}
\centerline{
\includegraphics[width=11cm]{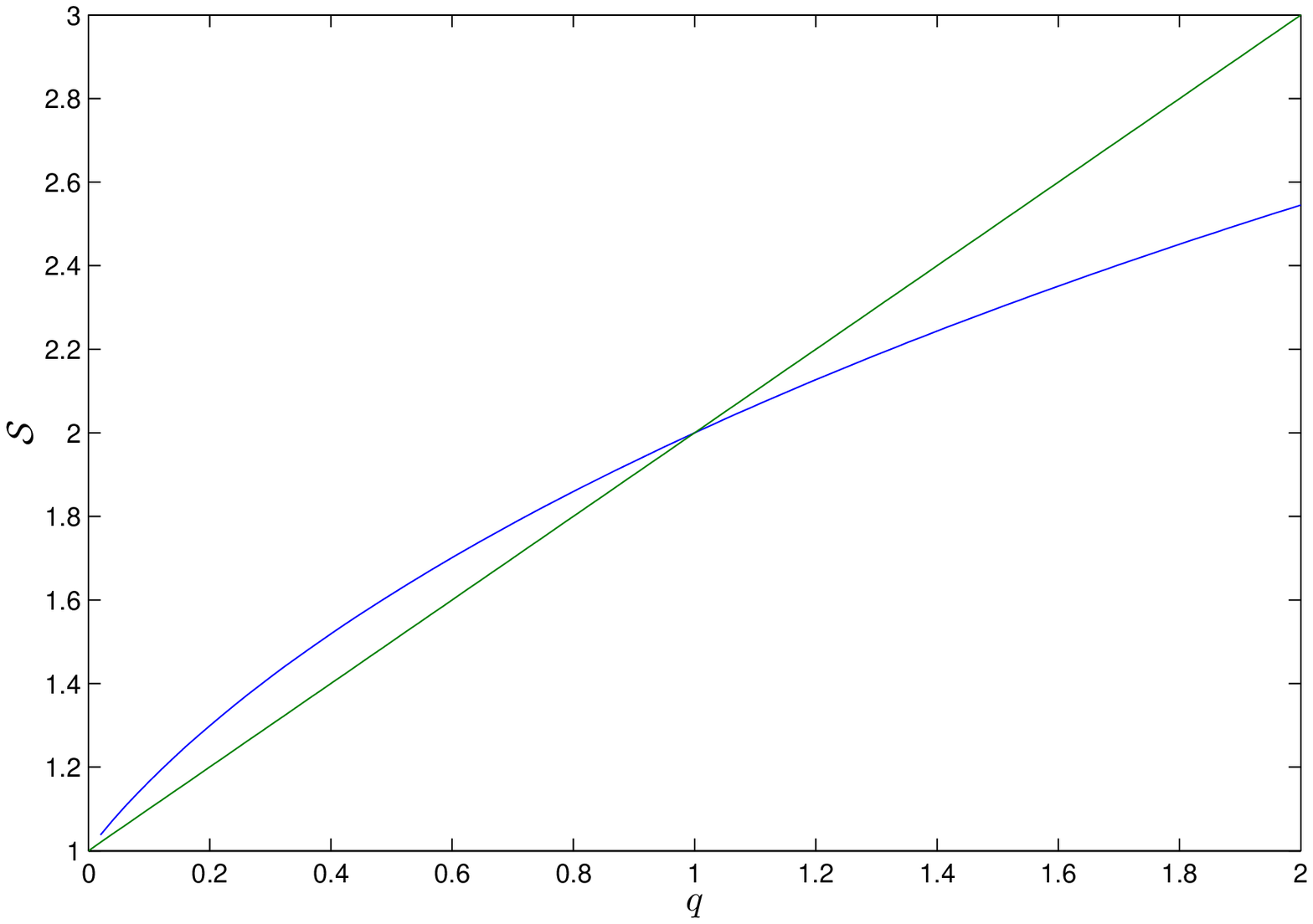}}
{\bf Fig.~8:} \small{(color online) Dependence of the subcriticality measures $\mathcal{S}(q)$
and $\mathcal{S}(q,q)$ given respectively by expression (\ref{yjuj}) (straight line) and (\ref{tu44j}) (concave curve).}
\end{quote}

\section{Conclusion}

Considering the class of multivariate self-excited Hawkes point processes,
we have presented the general theory of multivariate generating functions
to derive the number of events over all generations of various types that are triggered by a mother event
of a given type. This has allowed us to discuss in details the stability domains
of various systems, as a function of the topological structure of the mutual excitations
across different event types. In particular, we have studied
the case of symmetric mutual excitation abilities between events
of different types,  the case of just two different types of events, the case of
a one-dimensional chain of directed triggering in the space of event types
and the case of a one-dimensional chain in the space of event types with nearest-neighbor
triggering.  The main insight is that mutual triggering tends to provide
a significant extension of the stability (or subcritical) domain compared
with the case where event types are decoupled, that is, when an event of
a given type can only trigger events of the type.

\vskip 0.3cm
{\bf Acknowledgement}: We acknowledge financial support
from the ETH Competence Center "Coping with Crises in Complex
Socio-Economic Systems" (CCSS) through ETH Research Grant CH1-01-08-2
This work was also partially supported by ETH Research Grant ETH-31 10-3.

\end{document}